\documentclass[twocolumn,showpacs,preprintnumbers,prl,aps]{revtex4}

\usepackage{graphicx}
\usepackage{dcolumn}
\usepackage{bm}
\newcommand{\ber}{\begin{eqnarray}}
\newcommand{\eer}{\end{eqnarray}}
\newcommand{\bea}{\begin{equation}}
\newcommand{\eea}{\end{equation}}

\begin{document}

\title{Parametric internal waves in a compressible fluid}

\author{Kausik S. Das$^1$}
\email{kdas@physics.utoronto.ca}
\author{Stephen W. Morris$^1$}%
\author{A. Bhattacharyay$^2$}
\affiliation{
$^1$Department of Physics, University of Toronto,\\
60, St. George Street, Toronto, Ontario, Canada, M5S1A7.\\
$^2$IISER, Pune, Maharashtra 411008, India\\
}%

\begin{abstract}
We describe the effect of vibration on a confined volume of fluid which is density stratified due to its compressibility. We show that internal gravity-acoustic waves can be parametrically destabilized by the vibration.  The resulting instability is similar to the classic Faraday instability of surface waves, albeit modified by the compressible nature of the fluid.  It may be possible to observe experimentally near a gas-liquid critical point.

\end{abstract}


\maketitle

It is well known that when an incompressible, viscous fluid with a free surface undergoes vertical oscillations, standing waves can be parametrically excited which oscillate at half the frequency of the forcing~\cite{Faraday}. This {\it Faraday} instability occurs above a nonzero threshold acceleration which is related to the viscosity of the fluid~\cite{faraday_review}.  The patterns produced by the interactions of the waves are much-studied example of nonlinear pattern formation~\cite{patterns}.  Driven-dissipative parametric waves are described generically by a damped Mathieu equation~\cite{faraday_review,Satish}.  In this paper, we describe an application of this very general idea to internal waves in compressible, density stratified fluids.  We find that such a fluid can naturally support parametric waves and that such waves may be experimentally observable near a gas-liquid critical point.  This scenario has some important differences from previously studies cases of parametrically driven internal waves in incompressible fluids~\cite{BenielliJFM, Nature,schatz} because different modes of motion and different restoring forces arise in the compressible case.

The literature on internal waves in density stratified fluids is vast, owing to their relevance to atmospheric and oceanic processes~\cite{stratified}. Gravity waves, in which the restoring forces are due to the weight of the displaced fluid, propagate at a fixed angle to the direction of gravity which is determined by the wave frequency~\cite{igw}.  Experimentally, they may be parametrically excited by vertically vibrating the container~\cite{BenielliJFM}, and sidewall reflections can give rise to interesting self-reinforcing attractors~ \cite{Nature}.  In these experiments, the density stratification was produced compositionally, using dissolved salt.  Density gradients due to temperature gradients can also be excited parametrically by shaking the container, although the resulting waves may be mixed with other flows due to buoyancy driven convection~\cite{schatz}.  These previous studies of parametrically driven internal waves considered approximately incompressible fluids, that is, fluids in which the flow velocity associated with the wave motion was assumed to be negligible compared to the speed of sound. Here, we allow for the compressibility of the fluid, and thus allow for waves restored by non-hydrostatic pressure gradients, {\it i.e.} sound modes, in the fluid.

Even nominally incompressible fluids, like water, may experience tiny flows under vibration, due to their non-zero compressibility.  Numerous studies~\cite{Amin,Alexander2,Alexander1, Nelson,Homsy,Grassia,Bjarni} of shaken fluids have been undertaken both on Earth and in space, motivated by the non-ideal microgravity environment on orbiting spacecraft.  These effects, known collectively as {\it g-jitter}, are a significant barrier to using space-based platforms to study ideal gravity-free processes.  The experiments of Tryggvason {\it et. al.}~\cite{Bjarni} showed small coherent motion of neutrally buoyant particles in water under vibration in a sealed, isothermal cell.  These motions have the right scale to be caused by residual compressibility effects, and are difficult to explain any other way.  Such tiny flows are normally negligible in experiments, but were here seen superposed on the brownian motion and diffusion of microscopic particles in a fluid that would be motionless if it were perfectly incompressible.  Such flows could have an important influence on small scale mixing.

We have theoretically investigated a slab of compressible fluid subjected to both steady and oscillatory components of imposed acceleration.  We show that the oscillatory component can parametrically drive acoustic-gravity waves which are supported by the density stratification due to the constant component of the acceleration.  Our treatment generalizes a number of previous calculations~\cite{Livescu,Ribeyre,Puri}, which considered various limiting cases.

\section{Basic equations and scaling}

In this section we set up the hydrodynamic equations that describe the problem. We then consider the appropriate scaling of these equations for the rather unusual situation of a fluid which is both compressible and viscous.
The system consists of a laterally infinite layer of fluid, confined between two parallel surfaces on which we assume rigid boundary conditions.  The thickness of the layer is $d$.  The fluid is isentropic and compressible, with speed of sound $c$ and molecular viscosity $\eta$. The layer is subject to linear accelerations in the direction normal to the surfaces. For simplicity, we assume the combination of a constant and a sinusoidal acceleration, but the theory could easily be generalized to more complex linear accelerations.

Conservation of mass requires\cite{Acheson}
\begin{equation}
\nabla \cdot \vec{\bf u} = - \frac{1}{\rho} \frac{D \rho}{D t},
\label{div_u}
\end{equation}
where $D/D t = \partial/\partial t + (\vec{\bf u} \cdot \nabla)$ is the material derivative, $\rho$ is the density of the fluid and $\vec{\bf u}  =(u,v,w)$ is the fluid velocity.

Momentum conservation is described by the Navier-Stokes equations:
\begin{equation}
\rho \frac{D \vec{\bf u}}{D t}  = - \nabla p + \eta\nabla^2\vec{\bf u}-\frac{2}{3}\eta\nabla(\nabla\cdot\vec{\bf u})
- \rho  \hat{\bf z}\left[a_0+a_1 {\cos} (\omega t)\right].
\label{NSE}
\end{equation}
Here, $\hat{\bf z}$ is a unit vector normal to the surfaces.  $a_0$ is a constant acceleration, while $a_1$ is the amplitude of the sinusoidal driving acceleration.  On Earth, $a_0 = g$, and $\hat{\bf z}$ points upward.

The compressibility of the fluid implies that the pressure $p$ and the density $\rho$ are related by an equation of state such that
\begin{equation}
\frac{D p}{D t} = c^2 \frac{D \rho}{D t}, ~~~~~ c^2 = \bigg(\frac{\partial p}{\partial \rho}\bigg)_S,
\label{eq_state}
\end{equation}
where $c$ is the speed of sound and the subscript $S$ refers to constant entropy. To recover the incompressible limit, the speed of sound $c \rightarrow \infty$ and $D \rho/Dt \rightarrow 0$, such that their product remains finite. The appropriate boundary conditions on the velocity are the ``no slip" conditions, $\vec{\bf u}(z=\pm d/2) = 0$.

 In order to recast the problem in dimensionless form, we must choose scalings for the various quantities in Eq.~(\ref{div_u}-\ref{eq_state}).  Under physically reasonable vibration amplitudes, we expect the flows to be very weak, and therefore that the Reynolds number ${\cal R}$ will be small.  On the other hand, compressibility implies that the Mach number ${\cal M}$, while also very small, must be nonzero.  Accordingly, we choose a scaling such that the pressure gradient term balances the viscous term in Eq.~(\ref{NSE}), as would be expected for Stokes' flow.  We are lead to the following set of dimensionless variables, denoted by primes:
 \begin{eqnarray}
 (x^\prime, y^\prime, z^\prime) &=& (x/d, y/d, z/d),\nonumber \\
t^\prime &=& t\sqrt{\omega c/d}, \nonumber \\
\vec{\bf u}^\prime  &=& \vec{\bf u}/\sqrt{(\omega d c)},\nonumber \\
p^\prime &=& p\sqrt{d}/(\eta\sqrt{\omega c}),\\
\rho^\prime  &=& \rho c^2\sqrt{d}/(\eta\sqrt{\omega c}),\nonumber \\
a^\prime_0 &=& a_0/(\omega c),\nonumber \\
a^\prime_1 &=& a_1/(\omega c).\nonumber
 \end{eqnarray}
 In terms of these variables, omitting the primes, Eq.~(\ref{div_u}) is unchanged, and the other basic equations are
  \begin{eqnarray}
  {\cal M}\rho \frac{D \vec{\bf u}}{D t}  &=& -\nabla p +
 \nabla^2\vec{\bf u}-\frac{2}{3}\nabla(\nabla\cdot\vec{\bf u}) \nonumber\\&-& {\cal M} \rho  \hat{\bf z}\left[a_0+a_1 {\cos} (\sqrt{\cal M}t)\right]  \label{NSE_d}\\
 \frac{D p}{D t} &=& \frac{D \rho}{D t},  \label{eq_state_d}
 \end{eqnarray}
 where the Mach number ${\cal M} = \omega d / c$.   The Reynolds number under the same scaling is simply ${\cal R} = \omega d/c = {\cal M}$.  Thus, it is possible for both ${\cal M}$ and ${\cal R}$ to be small, while their ratio is $O(1)$.  This is a seldom used limit in fluid mechanics: that of Stokes' flow at nonzero Mach number. However, Eqs.~(\ref{div_u}) and (\ref{NSE_d}), (\ref{eq_state_d}) are still completely general, and no assumptions have been made about the size of ${\cal M}$.

 It is convenient to absorb a factor of ${\cal M}$ into the dimensionless acceleration parameters and define
 \begin{equation}
 A_0 = {\cal M} a^\prime_0 = d a_0/c^2,~~{\rm and}~~A_1 = {\cal M} a^\prime_1 = d a_1/c^2.
 \end{equation}
 We will show that these elementary equations contain sound modes and internal gravity wave modes which can be parametrically excited by the driving acceleration.

The base state corresponds to the case $A_1 = 0$, with $A_0$ arbitrary.  In this case, the fluid is static and there is a hydrostatic pressure distribution given by
\begin{equation}
\frac{d p}{d z}= - A_0 \rho.
\end{equation}

The corresponding density distribution is given by
\begin{equation}
\rho_0(z) = \rho_0(0) {e}^{- A_0 z}.
\end{equation}
where $\rho_0(0)$ is the density at $z = 0$. The base state is thus density stratified with scale height $A_0^{-1}$, which could be much larger than the thickness of the layer \cite{Das}. The typical order of magnitude of change in density of normal liquid like water in a 10 cm box in the ground based experiments is rather small and changes by only $O(10^{-4})\%$.

Density stratification due to  compressibility is fundamentally different than that due to salinity, the situation most often used in laboratory studies of internal waves~\cite{Nature}.  Nevertheless, such stratification is expected to become significant near the critical point of gases, where the speed of sound goes to zero and the compressibility diverges~\cite{SF6,crit_pt}.   In practice, the details of the equation of state of the gas are required to calculate the stratification.  Close to the critical point, the density profile will become quite nonlinear~\cite{crit_pt}.

\section {Linearised equations and numerical simulation}

We now consider the fluid under vibration and derive the equations for the growth or decay of small deviations about the ``base state".  We consider only a linear theory valid in the limit of small deviations.  This leads to a well-known equation for parametric oscillations.

To test the stability of the base state we consider the fluctuations $\delta \vec{\bf u}$, $\delta p$ and $\delta \rho$ of the velocity, pressure and density fields and linearise the equations of motion, Eqs.~(\ref{div_u}),   (\ref{NSE_d}) and  (\ref{eq_state_d}). Linearisation of Eq.~(\ref{div_u}) about the base state with $\rho=\rho_0$ and $\vec {\bf u}$=0 gives

\begin{equation}
\rho_0\nabla \cdot \delta \vec{\bf u} + \frac{\partial \delta \rho}{\partial t} + \delta w \frac{\partial \rho_0}{\partial z} = 0.
\label{div_u_pert}
\end{equation}
Considering the liquid to be irrotational, i.e., applying $\vec\nabla(\vec\nabla\cdot\delta\vec u)=\nabla^2\delta\vec u$, linearisation of Eq.~(\ref{NSE_d}) results in
\begin{eqnarray}
{\cal M} \rho_0 \frac{\partial \delta \vec{\bf u} }{\partial t} &=& -\nabla \delta p + \frac{1}{3}\nabla^2 \delta \vec{\bf u}\nonumber \\
&-& \delta \rho [A_0 +A_1 \cos(\sqrt{\cal M}t)]\hat{\bf z}\label{NSE_pert} \\
&-& \rho_0 A_1 \cos(\sqrt{\cal M}t) \hat{\bf z}.\nonumber
\end{eqnarray}
Finally, the equation of state reduces to
\begin{equation}
\frac{\partial \delta p}{\partial t}=\frac{\partial \delta \rho}{\partial t}
\label{eq_state_pert}
\end{equation}

 It is straightforward to eliminate $\delta \rho$ and $\delta p$ between Eqs.~(\ref{div_u_pert}), (\ref{NSE_pert}) and (\ref{eq_state_pert}). After some simplification and assuming $A_1$ and ${\cal M}$ are small so that the inhomogeneous forcing term can be neglected we obtain
\begin{eqnarray}
{\cal M}\frac{\partial^2 \delta \vec{\bf u}}{\partial t^2}-\frac{1}{3\rho_0}\nabla^2\frac{\partial\delta \vec{\bf u}}{\partial t} +A_0\nabla(\hat {z}\cdot\delta \vec{\bf u})=\nabla^2\delta \vec{\bf u}\nonumber\\+\hat{ z}\frac{A_1}{\rho_0}\Big[\rho_0\nabla\cdot\delta \vec{\bf u}+(\delta \vec{\bf u}\cdot\nabla)\rho_0\Big]{\rm \cos}(\sqrt{\cal M} t),
\label{parametric_eq}
\end{eqnarray}

If $A_0 = 0$, so that the constant component of the acceleration is zero, Eq.~(\ref{parametric_eq}) describes damped acoustic waves.  It can be shown from the dispersion relation for inviscid fluids that $\delta\vec{\bf u}$ and the wave vector $\vec{\bf k}$ are in the same direction and thus that such waves are purely longitudinal.  In the opposite limit,  when $A_1 = 0$ and the oscillatory component of the acceleration is zero,  Eq.~(\ref{parametric_eq}) reduces to a dynamical equation that supports mixed acoustic-gravity wave motion which is damped by viscosity.  Such mixed  waves are no longer completely longitudinal~\cite{Acheson}.

Eq.(\ref{parametric_eq}) was studied in two limited cases by \cite{Livescu},\cite{Ribeyre} and \cite{Puri}. In \cite{Livescu} and \cite{Ribeyre} the effects of compressibility on the early stages of Rayleigh-Taylor instability in non-viscous liquids was discussed. In \cite{Puri} it was shown that in absence of gravity for many liquids with low viscosity (e.g. air, water) Eq. (\ref{parametric_eq}) can be approximated by a damped wave equation. Moreover in the following analysis it is shown that this damping term for low viscous liquids has negligible effect on the stability of the stratified state in presence of jitter. However for brevity we kept this term in our analysis.

\begin{figure}[t]
\begin{center}
\includegraphics[width=8.5cm]{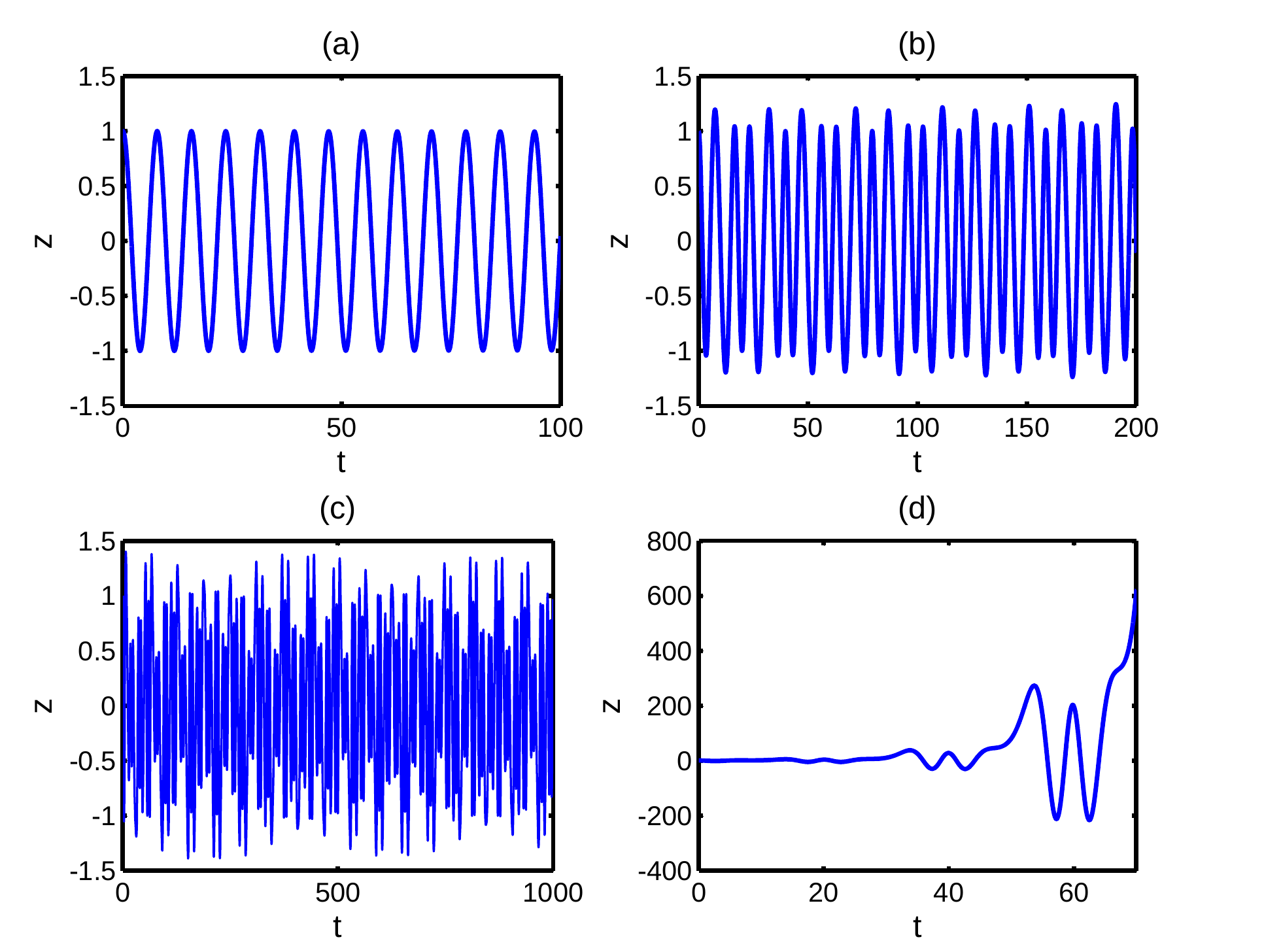}
\caption{Response of fluid particles with time. a) periodic response of fluid particles when density stratified base
state due to compressibility is perturbed in absence of jitter. $A_1=0, A_0=1.0, {\cal M}=0.3$.
b) amplitude is modulated by the parametric forcing term. $A_1=0.03, A_0=1.0,\alpha=0.1,{\cal M}=0.3$.
c) response for $A_1=0.05, A_0=1.0,\alpha=0.1, {\cal M}=0.3$, displacement of fluid particles remains bounded.
d) solution becomes unbounded when the value of $A_1$ is above the stability boundary specified by Eq.(\ref{ineq1}).
Linear theory breaks down in this case.}
\label{response}
\end{center}
\end{figure}

Eq.(\ref{parametric_eq}) is a parabolic, third-order partial differential
equation. Homogeneous form of Eq.(13) without jitter for
harmonic waves are solved by Stokes \cite{Stokes}
Stefan \cite{Stefan} and Rayleigh. Stokes' transient problem has two
types of solution : (a) solutions in the form of series \cite{Hanin,Norwood} or integrals \cite{Ludwig} and (b) closed-form approximations \cite{Blackstock}. Cobbold et al. \cite{Cobbold} has recently stated that a
difficulty arises with the approximate, closed-form type of solution : ``… approximate solutions to such problems
do not satisfy causality in the strict sense ….'' Many examples of breakdown of causality
for transient solutions of Stokes' equation may
be found in White \cite{White}. Jordan et.al \cite{Puri} and Blackstock\cite{Blackstock} have raised doubts over the validity of
Stokes' equation itself \cite{Buckingham}, with Jordan et. al \cite{Puri} stating that solutions for this problem ``...do not satisfy causality''. They along with others \cite{Blackstock,Norwood} have also observed that transient solutions of Stokes' equation
``are felt instantly'' throughout the entire fluid domain.
It has been established that an instantaneous response throughout the fluid
is unphysical, since an infinite wave speed is not permitted.
Weyman \cite{Weyman}, pointed out that this not only Stokes' equation but also other parabolic equations such as, the diffusion equation, is
commonly said to lead to an infinite speed of propagation
through the medium supporting the motion, even though
such behavior is entirely unphysical. Thus, it seems that the parabolic differential equations,even though derived correctly may lead to unphysical results. To avoid this paradox we followed Jordan et. al \cite{Puri} and approximated the damping term to convert the parabolic equation into a well-posed hyperbolic equation. For small $A_0$ and $A_1$ we assume the approximation \cite{Puri} is still valid and we have approximated Eq.(\ref{parametric_eq}) as
\ber
{\cal M}\frac{\partial^2 \delta \vec{\bf u}}{\partial t^2}+\frac{{\cal M}^2}{3\rho_0}\frac{\partial\delta \vec{\bf u}}{\partial t} +A_0\nabla(\hat {z}\cdot\delta \vec{\bf u})=\nabla^2\delta \vec{\bf u}\nonumber\\+\hat{ z}\frac{A_1}{\rho_0}\Big[\rho_0\nabla\cdot\delta \vec{\bf u}+(\delta \vec{\bf u}\cdot\nabla)\rho_0\Big]{\rm \cos}(\sqrt{\cal M} t),
\eer
 It follows that for liquids bounded by two parallel plates a solution of the form $ \delta\vec{\bf u} \sim\tilde{\vec{\bf u}}(t){\rm exp}\left(\alpha z+i(k_1x+k_2y)\right)$ can be sought.
Substituting this form and restricting ourselves to the long wavelength modes ($k_1,k_2\rightarrow 0$) we get an equation for the velocity component in the $\hat{z}$ direction i.e., $\delta \tilde{w}$ as
\begin{eqnarray}
{\cal M}\frac{{\rm d}^2 \tilde{w}(t)}{{\rm d} t^2}+\frac{{\cal M}^2}{3\rho_0}\frac{d \tilde{w}(t)}{d t}+\Big [\alpha(A_0-\alpha)\nonumber\\-A_1(\alpha-A_0){\cos}(\sqrt{\cal M}t)\Big]\tilde{w}(t)=0,
\label{ext_Mathieu}
\end{eqnarray}

where $\tilde{w}(t)$ is the time varying part of $\delta \tilde{w}$. This equation is a modified form of the Mathieu equation which describes parametric oscillations. It is evident that a parametric response to the driving frequency is possible for a range of parameters  ${\cal M},\; A_0,\; A_1$ and $\rho_0$, for a nonzero value of $\alpha$. We have solved Eq.~(\ref{ext_Mathieu}) numerically using a four point Runge-Kutta method and the results are shown in Fig.~\ref{response}. We find that in presence of constant acceleration, and in absence of shaking ($A_0 \neq 0$, $A_1=0$) the stratified fluid can undergo internal oscillations at the characteristic frequency $N$, where $N=\sqrt{\alpha(A_0-\alpha)}$.

 The amplitude of the response is modulated by the presence of shaking ($A_1\neq 0$).  The system becomes parametrically unstable if the amplitude of shaking exceeds that specified by the stability boundary derived analytically in the next section.

\section{Perturbation analysis}

Eq.~(\ref{ext_Mathieu}) with $A_1=0$, {\it i.e.}, in the absence of shaking, will always have oscillatory solutions for $A_0>\alpha$. But any viscosity, however small, will obviously cause the oscillations to damp out because the coefficient of the second term in Eq.~(\ref{ext_Mathieu}) is always positive. In what follows, we will show that such an oscillation of the density stratified fluid can grow under the action of periodic forcing or shaking.

Let us first rewrite Eq.~(\ref{ext_Mathieu}) as
\begin{equation}
\frac{{ d}^2 \tilde{w}(t)}{{ d} t^2}+K\frac{d \tilde{w}(t)}{d t}+\left(\beta+\epsilon~{\cos} (\sqrt{\cal M}t)\right)\tilde{w}(t)=0,
\label{Mathieu}
\end{equation}
where $K={{\cal M}}/{3\rho_0}$, $\beta={\alpha(A_0-\alpha)}/{{\cal M}}$ and $\epsilon={A_1(A_0-\alpha)}/{{\cal M}}$. For small $\epsilon $, it is possible to use very standard methods to do a perturbation analysis of Eq.~(\ref{Mathieu}).  This analysis finds the onset of linear instability and the regions where solutions of Eq.~(\ref{Mathieu}) become unbounded in the linearized theory. The only restriction imposed on the parameters is that $|\epsilon| \ll 1$. For small $|\epsilon|$, we expand $\beta$ and $\delta w$ in powers of $\epsilon$ as
\begin{equation}
\tilde{w} = w_0 + \epsilon w_1+{\epsilon}^2 w_2 + \cdots,
\end{equation}
and
\begin{equation}
\beta= \beta_0 + \epsilon \beta_1+\epsilon^2\beta_2 + \cdots.
\end{equation}
Putting the above expansions in Eq.(\ref{Mathieu}) we get the following equations at $O(1)$,  $O(\epsilon)$ and $O(\epsilon^2)$:
\begin{equation}
\frac{d^2 w_0}{dt^2}+K\frac{d w_0}{dt}+\beta_0 w_0 =0 ,
\label{leadingorder}
\end{equation}
\ber
\frac{d^2 w_1}{dt^2}+K\frac{d w_1}{dt}+\beta_0 w_1 = -\beta_1 w_0
\nonumber\\-\cos(\sqrt{\cal M}t)w_0 ,
\label{1storder}
\eer
\ber
\frac{d^2 w_2}{dt^2}+K\frac{d w_2}{dt}+\beta_0 w_2 = -\beta_2 w_0
-\beta_1 w_1 \nonumber\\-\cos(\sqrt{\cal M}t)w_1.
\label{2ndorder}
\eer

The solution of Eq.~(\ref{leadingorder}) is of the form $w_0 = b_0e^{\lambda t}$ with $\lambda = -K/2 \pm \sqrt{K^2-4\beta_0}/2 $, where $b_0$ is an arbitrary constant. For low viscosity liquids, $K$ is small and when $K^2<4\beta_0$ there will be an oscillatory component in the leading order solution.

When $w_0$ is put into Eq.~(\ref{1storder}), the $O(\epsilon)$ equation, the first term on the right hand side contains the secular term and has to be zero for the $O(\epsilon)$ solution to exist, according to Fredholm's condition~\cite{fredholm}. Thus, $\beta_1=0$, and the solution at $O(\epsilon)$ is
\ber
w_1 =b_1^0 \cos(\sqrt{\cal M}t)e^{\lambda t} + b_1^1 \sin(\sqrt{\cal M}t)e^{\lambda t} ~,
 \eer
 where $b_1^0=b_0/[1+{(K+2\lambda)}^2]$ and $b_1^1=-(K+2\lambda)b_1^0$.

Now, putting $w_0$ and $w_1$ into Eq.~(\ref{1storder}) and removing the secular term we find
$\beta_2=-({b_1^0}/{2})$
and $\beta$ calculated up to this order is
\ber
\beta=\beta_0-\epsilon^2\frac{b_0}{2[1+{(K+2\lambda)}^2]}
\eer
The $O(\epsilon^2)$ solution shows that the amplitude of the periodic response will be modulated by the parametric forcing term.

Next, we consider the transformation
\ber
\tilde{w}(t)=e^{-\frac{Kt}{2}}~w^{\prime}(t)
\eer
in order to cast the damped Mathieu equation, Eq.~({\ref{Mathieu}), into the form of a normal undamped Mathieu equation. The above transformation means that only those unbounded solutions of the Mathieu equation which grow faster than $e^{-{Kt}/{2}}$ will actually destabilize the oscillatory stratified state.
An analysis along standard lines using Floquet theory~\cite{jordan} reveals the first unstable region of the stratified state to occur when
\bea
(\beta-\frac{{\cal M}}{4})^2-\frac{1}{4}(\epsilon^2-{\cal M}K^2)\leq 0.
\eea
From the above inequality, the minimum shaking amplitude to induce unbounded motion turns out to be
\bea
A_1^2\geq\frac{{\cal M}^5}{9\rho_0^2{(\alpha-A_0)}^2}+\frac{4[\alpha(A_0-\alpha)-({{\cal M}^2}/{4})]^2}{{(\alpha-A_0)}^2}.
\label{ineq1}
\eea

\begin{figure}[ht]
\begin{center}
\includegraphics[width=8.5cm,angle=0]{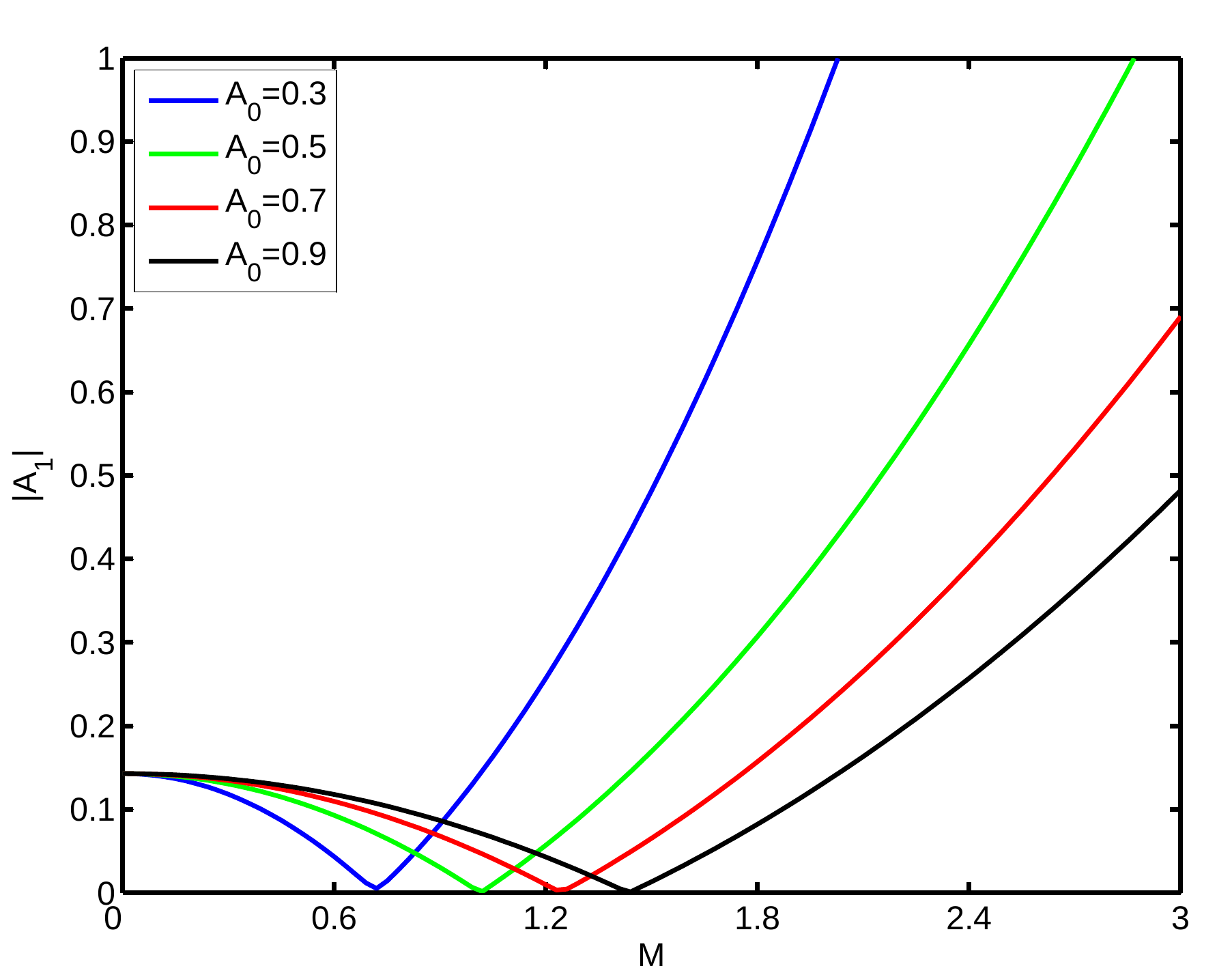}
\caption{The stability regions in the forcing amplitude $A_1$ {\it vs.} Mach number ${\cal M}$ plane, for a fluid with negligible viscosity.  The region above the curves is parametrically unstable, and the internal waves will grow to become nonlinear.  The sharp minimum means that forcing with infinitesimal amplitude may destabilize the system at those parameter regimes.}

\label{stability_boundaries}
\end{center}
\end{figure}

We can rewrite this result in dimensional form as
\bea
a_1^2\geq\frac{\omega^6\eta^2}{9\rho_0^2{(\tilde{\alpha}c^2-a_0)}^2}+\frac{4c^4[\tilde{\alpha} ( a_0-\tilde{\alpha} c^2)-({\omega^2}/{4})]^2}{{(\tilde{\alpha}c^2-a_0)}^2},
\label{ineq2}
\eea

where $\tilde{\alpha}=\alpha /d$ and $\tilde{k}=k/d$. For many liquids like water the first term on the r.h.s. of
Eq.~(\ref{ineq2}) is negligible since kinematic viscosity $\nu=\eta/{\rho_0}\sim 10^{-6}$~m/s is quite small. The second term then dominates and determines the minimum forcing amplitude.  The corresponding forcing frequency is given by $\omega_0^2=4\tilde{\alpha} (a_0-\tilde{\alpha} c^2)$. Typically, there will exist a narrow band of frequencies for which the system is unstable just above this minimum forcing amplitude.

In Fig.~\ref{stability_boundaries}, we plot the stability threshold for $A_1$ against ${\cal M}$ for different values of $A_0$, assuming a fluid of negligible kinematic viscosity.
We find that driving frequencies close to a subharmonic resonance with the characteristic frequency $N$ most easily drive
the internal waves.  It is also interesting that waves can be driven parametrically with modest values of $A_1$, even at
frequencies well below $N/2$.
%
%


It is important to note that standard results for the damped Mathieu equation~\cite{jordan} indicate that  higher order resonances will be encountered for higher positive values of the quantity $\sigma = \beta-({K^2}/{4})$, where
\begin{equation}
\sigma=\frac{\alpha(A_0-\alpha)}{{\cal M}}-\frac{{\cal M}^2}{12\rho_0^2}.
\label{last}
\end{equation}
The second term on the r.h.s. of Eq.~(\ref{last}) is negligible for low viscosity fluids.  General stability diagrams which show the higher order resonances in the $\sigma -K$ plane are beyond the reach of perturbation theory and must be obtained numerically~\cite{jordan}.  These higher order regions have considerable spread in the parameter space, and thus for a higher forcing amplitude it is more probable that the system will be in a region of unbounded solutions.

In Eq.~(\ref{last}), the first term is positive when $A_0>\alpha$ and can slowly be increased if ${\cal M}$ is gradually decreased.  This suggests a possible experimental protocol to fix the driving frequency and acceleration and slowly increase the temperature of the gas away from near the gas-liquid critical point, where the speed of sound becomes small.  This has the effect of decreasing ${\cal M}$.  One could then in principle successively observe the various higher order resonances which are characteristic of the Mathieu equation.  A detailed calculation using real gas equations of state, combined with numerical solutions of Eq.~(\ref{ext_Mathieu}) would be necessary to make specific predictions for experiments, however.

\section {Conclusion}

We have investigated the effect of vibration on a compressible fluid which may also be subject to a constant acceleration. The response of the fluid can show both harmonic and subharmonic components, as well as possible higher order resonances.  The harmonic response probably explains the coherent oscillations observed by Tryggvasson {\it et al}~\cite{Bjarni} in a sealed water cell.   The oscillatory acceleration can interact with the stratification produced by the constant acceleration to produce parametrically excited internal waves.  For fluids of small viscosity, the threshold for this instability can be very small if the frequency is selected to be close to a subharmonic of the Brunt-V$\ddot{\rm a}$is$\ddot{\rm a}$l$\ddot{\rm a}$ frequency of the stratification.  All these effects may be observable in fluids near their critical points, where compressible effect become very large~\cite{Das}.

\begin{acknowledgments}
KSD acknowledges the Canadian Space Agency for providing financial support through Contract No. 9F007-046036/001/ST and Bjarni Tryggvason for valuable comments and discussions.
\end{acknowledgments}

\end{document}